\documentclass[twocolumn,superscriptaddress,
showpacs,preprintnumbers,amsmath,amssymb]{revtex4}
\usepackage{graphicx}% Include figure files

\newcommand{\bra}[1]{{\left\langle #1 \right|}}
\newcommand{\ket}[1]{{\left| #1 \right\rangle}}

\newcommand{\T}{\mbox{$\mathrm{tr}$}}

\begin{document}

%%%%%%%%%%%%%%%%%%%%%%%%%%%%%%%%%%%%%%%%%%%%%%%%%%%%%%%%%%%%%%%%%%%%%%%%%%
%                                                                        %
%                                 Title                                  %
%                                                                        %
%%%%%%%%%%%%%%%%%%%%%%%%%%%%%%%%%%%%%%%%%%%%%%%%%%%%%%%%%%%%%%%%%%%%%%%%%%
\title{Monogamy of entanglement and teleportation capability}

\author{Soojoon Lee}%\email{level@khu.ac.kr}
\affiliation{
 Department of Mathematics and Research Institute for Basic Sciences,
 Kyung Hee University, Seoul 130-701, Korea
}
\author{Jungjoon Park}%\email{ilvolo@uos.ac.kr}
\affiliation{
 Division of Liberal Art in Mathematics,
 University of Seoul, Seoul 130-743, Korea
}

\date{\today}

%%%%%%%%%%%%%%%%%%%%%%%%%%%%%%%%%%%%%%%%%%%%%%%%%%%%%%%%%%%%%%%%%%%%%%%%%%
%                                                                        %
%                              Abstract                                  %
%                                                                        %
%%%%%%%%%%%%%%%%%%%%%%%%%%%%%%%%%%%%%%%%%%%%%%%%%%%%%%%%%%%%%%%%%%%%%%%%%%
\begin{abstract}
The monogamy inequality in terms of the concurrence,
called the Coffman-Kundu-Wootters inequality~[Phys. Rev. A {\bf 61}, 052306 (2000)],
and its generalization~[T.J.~Osborne and F.~Verstraete, Phys. Rev. Lett. {\bf 96}, 220503 (2006)]
hold on general $n$-qubit states including mixed ones.
In this paper,
we consider the monogamy inequalities
in terms of the fully entangled fraction and the teleportation fidelity.
We show that the monogamy inequalities do not hold on general mixed states,
while the inequalities hold on $n$-qubit pure states.
\end{abstract}

%\keywords{}
\maketitle

%%%%%%%%%%%%%%%%%%%%%%%%%%%%%%%%%%%%%%%%%%%%%%%%%%%%%%%%%%%%%%%%%%%%%%
%%%                                                                %%%
%%%                         Introduction                           %%%
%%%                                                                %%%
%%%%%%%%%%%%%%%%%%%%%%%%%%%%%%%%%%%%%%%%%%%%%%%%%%%%%%%%%%%%%%%%%%%%%%
%\section{Introduction}
Entanglement is a crucial one of several quantum mechanical phenomena,
which provide us with various quantum information processing
such as quantum computation superior to any classical computation
and perfectly secure quantum communication.
On this account,
over the last three decades,
a lot of research works on entanglement have been significantly developed.
However, there still exist quite a few unsolved problems
and interesting aspects related to entanglement,
especially multipartite entanglement.

In order to understand entanglement more effectively,
quantum information scientists have considered
the so-called {\it entanglement measure}
as a tool to quantify the degree of entanglement.
Among several entanglement measures,
there is a simple form of entanglement measure,
called the {\it concurrence}~\cite{Wootters},
which is defined as follows:
For a given pure state $\ket{\phi}_{12}$ in $2\otimes d$ quantum system ($d\ge 2$),
the concurrence $\mathcal{C}$ is defined as
$\mathcal{C}(\ket{\phi}_{12}\bra{\phi})=\sqrt{2(1-\T\rho_1^2)}=2\sqrt{\det\rho_1}$,
where $\rho_1=\T_2\ket{\phi}_{12}\bra{\phi}$,
and for a given mixed state $\rho_{12}$,
\begin{equation}
\mathcal{C}(\rho_{12})=\min \sum_{k} p_k \mathcal{C}(\ket{\phi_k}_{12}),
\label{eq:concurrence}
\end{equation}
where the minimum is taken over its all possible decompositions,
$\rho_{12}=\sum_k p_k \ket{\phi_k}_{12}\bra{\phi_k}$.

The concurrence has been considered
as one of the most important measures of entanglement,
since the computable formula for the entanglement of formation
in the two-qubit system can be derived from
the explicit formula for the concurrence in the two-qubit system, and
various properties about entanglement can be seen by virtue of the concurrence.

One of those interesting properties about multipartite entanglement is
the {\it monogamy of entanglement}, which
can be seen by the monogamy inequality in terms of
the concurrence as follows:
For a given $n$-qubit pure states $\ket{\Psi}_{12\cdots n}$
and its reduced density operators $\rho_{1j}$,
\begin{equation}
\mathcal{C}_{1(2\cdots n)}^2\ge\mathcal{C}_{12}^2+\mathcal{C}_{13}^2+\cdots+\mathcal{C}_{1n}^2,
\label{eq:CKWn}
\end{equation}
where $\mathcal{C}_{1(2\cdots n)}=\mathcal{C}(\ket{\Psi}_{1(2\cdots n)}\bra{\Psi})$
and $\mathcal{C}_{1j}=\mathcal{C}(\rho_{1j})$.
The inequality is
called the Coffman-Kundu-Wootters inequality~\cite{CKW} when $n=3$,
%which was shown by Coffman {\it et al.}
and
its generalization to multiqubit states
was recently proved by Osborne and Verstraete~\cite{OV} as well.
%%%
%%% New 1
%%%
The monogamy inequality tells us that
if one particle $P$ in a multipartite quantum system is entangled with another particle $Q$
as much as entanglement between $P$ and the other particles,
then $P$ cannot be entangled with any other particles except $Q$.

There are several useful applications of entanglement,
among which a practical one is teleportation~\cite{BBCJPW}.
Since teleportation is to send quantum information
through a classical channel assisted by entanglement,
the teleportation capability essentially depends on
the degree of the entanglement to assist the classical channel.
Thus, we need to review the relation between the entanglement
and the teleportation capability.

We now consider a natural quantity related to the teleportation capability,
called the {\it teleportation fidelity}~\cite{Popescu},
which is defined by
\begin{equation}
f(\Lambda_{\rho})=\int d\xi \bra{\xi}\Lambda_{\rho}(\ket{\xi}\bra{\xi})\ket{\xi},
\label{eq:teleportation_fidelity}
\end{equation}
where $\Lambda_\rho$ is a given teleportation scheme over a state $\rho$,
and the integral is performed
with respect to the uniform distribution $d\xi$ over all one-qubit pure states.
Let $F(\rho)$ be
the {\it fully entangled fraction}~\cite{BBPSSW,Horodeckis1,Horodeckis2,BadziagHorodeckis} of $\rho$
defined as
\begin{equation}
F(\rho)=\max\bra{e}\rho\ket{e},
\label{eq:FEF}
\end{equation}
where the maximum is over all maximally entangled states $\ket{e}$.
Then it has been shown~\cite{Horodeckis2,BadziagHorodeckis} that
the maximal fidelity achievable
from a given state $\rho$ in $2\otimes d$ quantum system
is
\begin{equation}
f(\Lambda_{\rho})=\frac{2F(\rho)+1}{3},
\label{eq:2relation}
\end{equation}
where $\Lambda_{\rho}$ is the standard teleportation scheme over $\rho$
to attain the maximal fidelity.

Let $\ket{\phi}$ be a pure state in $2\otimes d$ quantum system with $d\ge 2$.
Then $\ket{\phi}$ can be expressed as
\begin{equation}
\ket{\phi}=\sqrt{\alpha}\ket{a_0 b_0}+\sqrt{\beta}\ket{a_1 b_1}
\label{eq:Schimidt}
\end{equation}
by the Schmidt decomposition theorem,
where $\sqrt{\alpha}$ and $\sqrt{\beta}$ are the Schmidt coefficients,
and $\{\ket{a_j}\}$ and $\{\ket{b_j}\}$ are the set of orthonormal vectors
in $\mathbb{C}^2$ and $\mathbb{C}^d$, respectively.
Thus, we can readily obtain that
\begin{eqnarray}
F(\ket{\phi}\bra{\phi})&=&1/2+\sqrt{\alpha\beta},\nonumber\\
\mathcal{C}(\ket{\phi}\bra{\phi})&=&2\sqrt{\alpha\beta},
\label{eq:F_C}
\end{eqnarray}
and hence
\begin{equation}
\mathcal{C}(\ket{\phi}\bra{\phi})=2F(\ket{\phi}\bra{\phi})-1
=3f(\Lambda_{\ket{\phi}\bra{\phi}})-2
\label{eq:C_F}
\end{equation}
for any pure state $\ket{\phi}$ in $2\otimes d$ quantum systems.

Let $\rho=\sum_k p_k \ket{\psi_k}\bra{\psi_k}$ be an optimal decomposition
of a state $\rho$ in $2\otimes d$ quantum system
with respect to the concurrence.
Then since $F$ is convex, it follows that
\begin{eqnarray}
\mathcal{C}(\rho)
&=&\sum_k p_k \mathcal{C}\left(\ket{\psi_k}\bra{\psi_k}\right)\nonumber \\
&=&2\sum_k p_k  F\left(\ket{\psi_{k}}\bra{\psi_{k}}\right)-1 \nonumber \\
&\ge&2F\left(\rho\right)-1  \nonumber \\
&=&3f\left(\Lambda_{\rho}\right)-2,
\label{eq:C_F_mixed}
\end{eqnarray}
by Eqs.~(\ref{eq:2relation}) and (\ref{eq:C_F}).
Therefore, by Eqs.~(\ref{eq:CKWn}), (\ref{eq:C_F}), and (\ref{eq:C_F_mixed}),
we have the following monogamy inequalities in terms of
the fully entangled fraction and the teleportation fidelity
on $n$-qubit pure states, respectively.
\begin{eqnarray}
(2F_{1(2\cdots n)}-1)^2&\ge&(2F_{12}-1)^2%+(2F_{13}-1)^2
+\cdots+(2F_{1n}-1)^2,
\nonumber \\
(3f_{1(2\cdots n)}-2)^2&\ge&(3f_{12}-2)^2%+(3f_{13}-2)^2
+\cdots+(3f_{1n}-2)^2,
\nonumber \\
\label{eq:monogamy_ineqs}
\end{eqnarray}
where
\begin{eqnarray}
F_{1(2\cdots n)}&=&F(\ket{\Psi}_{1(2\cdots n)}\bra{\Psi}),
\nonumber \\
f_{1(2\cdots n)}&=&f(\Lambda_{\ket{\Psi}_{1(2\cdots n)}\bra{\Psi}}),
\nonumber \\
F_{1j}&=&\max\{F(\rho_{1j}),1/2\},
\nonumber \\
f_{1j}&=&\max\{f(\Lambda_{\rho_{1j}}),2/3\}.
\label{eq:F_f}
\end{eqnarray}
%%%
%%% New 2
%%%
We here note that $F(\rho)>1/2$ (or $f(\Lambda_{\rho})>2/3$)
if and only if $\rho$ is said to be useful for teleportation,
since it has been shown that the classical teleportation
can have at most $F=1/2$ (or $f=2/3$)~\cite{Popescu,Horodeckis1}.
This is the reason why
the maxima with 1/2 and 2/3 are used
in the definitions of $F_{1j}$ and $f_{1j}$ in~(\ref{eq:F_f}), respectively.

Remark that the monogamy inequality~(\ref{eq:CKWn}) in terms of the concurrence also holds
on $n$-qubit mixed states,
as one can see in the proof of the generalization of the Coffman-Kundu-Wootters inequality~\cite{OV}.
Then one could naturally ask
whether the monogamy inequalities in terms of the fully entangled fraction
and the teleportation fidelity hold on $n$-qubit mixed states or not.
%%%
%%% New 3
%%%
If the monogamy inequalities would be satisfied on mixed states as well as pure states,
then one could know that
faithful teleportation cannot be freely performed between any pair of particles
in a given multiqubit entangled state.
%%%
In this paper, we give the answer to the question.

%%%
%%%    organization of this paper
%%%
%This paper is organized as follows.

%%%%%%%%%%%%%%%%%%%%%%%%%%%%%%%%%%%%%%%%%%%%%%%%%%%%%%%%%%%%%%%%%%%%%%
%%%                                                                %%%
%%%                     Fundamental Properties                     %%%
%%%                                                                %%%
%%%%%%%%%%%%%%%%%%%%%%%%%%%%%%%%%%%%%%%%%%%%%%%%%%%%%%%%%%%%%%%%%%%%%%
%\section{Fundamental Properties}
In order to answer to the question,
we first need to know some properties of the fully entangled fraction.
Let $\ket{\phi^{\pm}}$ and $\ket{\psi^{\pm}}$ be the Bell states
in $2\otimes 2$ quantum system,
that is,
\begin{eqnarray}
\ket{\phi^{\pm}}&=&\frac{1}{\sqrt{2}}\left(\ket{00}\pm\ket{11}\right),
\nonumber \\
\ket{\psi^{\pm}}&=&\frac{1}{\sqrt{2}}\left(\ket{01}\pm\ket{10}\right),
\label{eq:Bell_states}
\end{eqnarray}
and let
\begin{eqnarray}
\ket{\tilde{\phi}^{\pm}}&=&\frac{1}{\sqrt{2}}\left(\ket{000}\pm\ket{101}\right),
\nonumber \\
\ket{\tilde{\psi}^{\pm}}&=&\frac{1}{\sqrt{2}}\left(\ket{001}\pm\ket{100}\right)
\label{eq:24Bell_states}
\end{eqnarray}
be the states in $2\otimes 2\otimes 2$ quantum system.
Then $\ket{\tilde{\phi}^{\pm}}$ and $\ket{\tilde{\psi}^{\pm}}$ can be regarded
as the states in $2\otimes 4$ quantum system as well.
We note that if $A$ is any $2\times 2$ matrix and $B$ is any $4\times 4$ matrix
then, by tedious but straightforward calculation,
we can show that
\begin{equation}
\left(A\otimes B\right)\ket{\tilde{\phi}^+}
=\left(I\otimes B\tilde{A}^T\right)\ket{\tilde{\phi}^+},
\label{eq:AB_IBA}
\end{equation}
where $\tilde{A}$ is the $4\times 4$ matrix satisfying
$\tilde{A}\ket{j}=A\ket{j}$ for $j=0,1$ and
$\tilde{A}\ket{j}=\ket{j}$ for $j=2,3$,
that is,
\begin{equation}
\tilde{A}=
\begin{pmatrix}
A & 0 \\
0 & I
\end{pmatrix}.
\label{eq:tilde_A}
\end{equation}
Thus, we can obtain the following equalities
for the fully entangled fraction of states in $2\otimes 4$ quantum system:
For a state $\rho$ in $2\otimes 4$ quantum system,
\begin{eqnarray}
F(\rho)&=& \max_{\ket{e}}\bra{e}\rho\ket{e} \nonumber \\
&=& \max_{U, V} \bra{\tilde{\phi}^+}
\left(U^\dagger\otimes V^\dagger\right)\rho\left(U\otimes V\right)
\ket{\tilde{\phi}^+}
\label{eq:FEF24UV} \\
&=& \max_{V} \bra{\tilde{\phi}^+}
\left(I\otimes V^\dagger\right)\rho\left(I\otimes V\right)
\ket{\tilde{\phi}^+},
\label{eq:FEF24V}
\end{eqnarray}
where the maximum in Eq.~(\ref{eq:FEF24UV})
is over all $2\times 2$ unitary matrices $U$
and all $4\times 4$ unitary matrices $V$,
and the maximum in Eq.~(\ref{eq:FEF24V})
is over all $4\times 4$ unitary matrices $V$.

%%%%%%%%%%%%%%%%%%%%%%%%%%%%%%%%%%%%%%%%%%%%%%%%%%%%%%%%%%%%%%%%%%%%%%
%%%                                                                %%%
%%%                        Couterexample                           %%%
%%%                                                                %%%
%%%%%%%%%%%%%%%%%%%%%%%%%%%%%%%%%%%%%%%%%%%%%%%%%%%%%%%%%%%%%%%%%%%%%%
%\section{Counterexample}
We now take account of a 2-parameter class of states in $2\otimes d$ quantum system
proposed by Chi and Lee~\cite{CL},
which can be described as a generalization of the Werner state~\cite{Werner}
to $2\otimes d$ quantum system:
\begin{eqnarray}
\rho_{(\alpha,\gamma)}
&=&\alpha \sum_{i=0}^1\sum_{j=2}^{d-1}\ket{ij}\bra{ij} \nonumber \\
&&+\beta \left(\ket{\tilde{\phi}^+}\bra{\tilde{\phi}^+}
+\ket{\tilde{\phi}^-}\bra{\tilde{\phi}^-}
+\ket{\tilde{\psi}^+}\bra{\tilde{\psi}^+}\right)
\nonumber \\
&&+\gamma \ket{\tilde{\psi}^-}\bra{\tilde{\psi}^-},
\label{eq:2parameter}
\end{eqnarray}
where $0\le \alpha, \beta, \gamma \le 1$ and $2(d-2)\alpha+3\beta+\gamma=1$.
For convenience, we here deal with the case of $d=4$ only.

Note that the 4-dimensional standard basis vectors,
$\ket{0}$, $\ket{1}$, $\ket{2}$, and $\ket{3}$,
can be clearly identified with
the standard basis vectors in $2\otimes 2$ quantum system,
$\ket{00}$, $\ket{01}$, $\ket{10}$, and $\ket{11}$, respectively.
Thus, $\ket{ij}$ for $i=0,1$ and $j=2,3$
can be also identified with $\ket{i1j}$ $i=0,1$ and $j=0,1$, respectively.

If we take $\alpha=\beta$
then $\rho_{(\alpha,\gamma)}$ in Eq.~(\ref{eq:2parameter}) can be rewritten as
\begin{eqnarray}
\rho_{(\alpha,\gamma)}
&=&\alpha \sum_{i=0}^1\sum_{j=0}^{1}\ket{i1j}\bra{i1j} \nonumber \\
&&+\alpha \left(\ket{\tilde{\phi}^+}\bra{\tilde{\phi}^+}
+\ket{\tilde{\phi}^-}\bra{\tilde{\phi}^-}
+\ket{\tilde{\psi}^+}\bra{\tilde{\psi}^+}\right)
\nonumber \\
&&+\gamma \ket{\tilde{\psi}^-}\bra{\tilde{\psi}^-}.
\label{eq:2parameter01}
\end{eqnarray}
Since $\alpha=(1-\gamma)/7$,
we may let $\sigma_\gamma^{123}$ be the 1-parameter class of states
such that
$\sigma_\gamma^{123}\equiv\rho_{(\alpha,\gamma)}$ in Eq.~(\ref{eq:2parameter01}).

We assume that $\gamma > \alpha$.
Then, by Eq.~(\ref{eq:FEF24V}),
we can calculate the fully entangled fraction of $\sigma_\gamma^{123}$
with respect to $2\otimes 4$ quantum system,
$F(\sigma_\gamma^{1(23)})$, as follows.
\begin{eqnarray}
F_{1(23)}
&\equiv& F(\sigma_\gamma^{1(23)}) \nonumber \\
&=& \max_{U} \bra{\tilde{\phi}^+}
\left(I\otimes U^\dagger\right)\sigma_\gamma^{1(23)}\left(I\otimes U\right)
\ket{\tilde{\phi}^+}
\nonumber \\
&=& \frac{1}{2} \max_U
\left(2\alpha+\frac{\gamma-\alpha}{2}\left|\bra{00}U\ket{01}-\bra{01}U\ket{00}\right|^2\right)
\nonumber \\
&=& \gamma,
\label{eq:F_1_23}
\end{eqnarray}
where the maximum in Eq.~(\ref{eq:F_1_23})
is over all $4\times 4$ unitary matrices $U$,
and the third equality can be directly obtained
by routine calculations.
Thus, we also have
$f_{1(23)}\equiv f(\Lambda_{\sigma_\gamma^{1(23)}}) =(2\gamma+1)/3$.

We now take the reduced density operator $\sigma_\gamma^{13}$ of $\sigma_\gamma^{123}$ into account.
By tracing out the second qubit system,
it follows that
\begin{eqnarray}
\sigma_\gamma^{13}
&=&\alpha \sum_{i=0}^1\sum_{j=0}^{1}\ket{ij}\bra{ij} \nonumber \\
&&+\alpha \left(\ket{{\phi}^+}\bra{{\phi}^+}
+\ket{{\phi}^-}\bra{{\phi}^-}
+\ket{{\psi}^+}\bra{{\psi}^+}\right)
\nonumber \\
&&+\gamma \ket{{\psi}^-}\bra{{\psi}^-}.
\label{eq:sigma_12}
\end{eqnarray}
Since
\begin{eqnarray}
\ket{{\phi}^+}\bra{{\phi}^+}+\ket{{\phi}^-}\bra{{\phi}^-}
&=& \ket{00}\bra{00} + \ket{11}\bra{11} \nonumber \\
\ket{{\psi}^+}\bra{{\psi}^+}+\ket{{\psi}^-}\bra{{\psi}^-}
&=& \ket{01}\bra{01} + \ket{10}\bra{10},
\label{eq:Bell_separable}
\end{eqnarray}
$\sigma_\gamma^{13}$ can be clearly redescribed as
\begin{eqnarray}
\sigma_\gamma^{13}
&=&2\alpha \left(\ket{{\phi}^+}\bra{{\phi}^+}
+\ket{{\phi}^-}\bra{{\phi}^-}
+\ket{{\psi}^+}\bra{{\psi}^+}\right)
\nonumber \\
&&+(\alpha+\gamma)\ket{{\psi}^-}\bra{{\psi}^-}.
\label{eq:sigma_12_simple}
\end{eqnarray}
Then, by the explicit formula for the fully entangled fraction of 2-qubit states,
we clearly have
\begin{eqnarray}
F_{13}\equiv F(\sigma_\gamma^{13}) = \alpha+\gamma = \frac{6\gamma+1}{7},
\nonumber \\
f_{13}\equiv f(\Lambda_{\sigma_\gamma^{13}}) = \frac{4\gamma+3}{7},
\label{eq:Ff_13}
\end{eqnarray}
and hence if $\gamma<1$ then $F_{1(23)} < F_{13}$ and $f_{1(23)} < f_{13}$.
Therefore, the monogamy inequality in terms of the fully entangled fraction,
\begin{eqnarray}
(2F_{1(23)}-1)^2 &\ge& (2F_{12}-1)^2+(2F_{13}-1)^2, %\nonumber \\
%&\ge& (2F_{13}-1)^2,
\label{eq:monogamy_ineqFEF3}
\end{eqnarray}
is violated
since $(2F_{1(23)}-1)^2 < (2F_{13}-1)^2$ when $\gamma<1$,
and so the inequality in terms of the teleportation fidelity,
\begin{equation}
(3f_{1(23)}-2)^2 \ge (3f_{12}-2)^2+(3f_{13}-2)^2,
\label{eq:monogamy_ineqTF3}
\end{equation}
does not hold in general.
Furthermore, this also shows that
there exist states not satisfying the equality in the inequality (\ref{eq:C_F_mixed}),
that is, if $\gamma<1$ then
\begin{eqnarray}
\mathcal{C}(\sigma_\gamma^{1(23)})&>& 2F(\sigma_\gamma^{1(23)})-1,
\nonumber \\
\mathcal{C}(\sigma_\gamma^{1(23)})&>& 3f(\Lambda_{\sigma_\gamma^{1(23)}})-2.
\label{eq:not_equal}
\end{eqnarray}

%%%%%%%%%%%%%%%%%%%%%%%%%%%%%%%%%%%%%%%%%%%%%%%%%%%%%%%%%%%%%%%%%%%%%%
%%%                                                                %%%
%%%                        Conclusion                              %%%
%%%                                                                %%%
%%%%%%%%%%%%%%%%%%%%%%%%%%%%%%%%%%%%%%%%%%%%%%%%%%%%%%%%%%%%%%%%%%%%%%
%\section{Conclusion}\label{sec:4}
In conclusion,
we have considered the monogamy of entanglement given by the monogamy inequalities
in terms of the fully entangled fraction and teleportation fidelity.
We have proved that
while the monogamy inequalities holds on $n$-qubit pure states,
the inequalities does not hold on general mixed states.

The violation of the monogamy inequalities has been shown
by exhibiting a one-parameter class of three-qubit states
whose fully entangled fraction or teleportation fidelity
are less than those quantities for their reduced density operators to a 2-qubit subsystem.
Therefore, the existence of this class of the states also tells us that
the fully entangled fraction and the teleportation fidelity
are meaningful quantities to represent the teleportation capability,
but cannot be suitable entanglement measures to quantify the degree of entanglement.

%\newpage
%%%%%%%%%%%%%%%%%%%%%%%%%%%%%%%%%%%%%%%%%%%%%%%%%%%%%%%%%%%%%%%%%%%%%%
%%%                                                                %%%
%%%                       Acknowledgements                         %%%
%%%                                                                %%%
%%%%%%%%%%%%%%%%%%%%%%%%%%%%%%%%%%%%%%%%%%%%%%%%%%%%%%%%%%%%%%%%%%%%%%
%\acknowledgments{
This research was supported by the Kyung Hee University
Research Fund in 2008 (KHU-20080578).
%}
%\newpage

\end{document}